\documentclass[aps,reprint,superscriptaddress,nofootinbib,longbibliography,floatfix]{revtex4-2}

\usepackage[T1]{fontenc}
\usepackage[utf8]{inputenc}
\usepackage{graphicx}
\usepackage{amsmath}
\usepackage{amssymb}
\usepackage{amsbsy}
\usepackage{bm}
\usepackage{xcolor}
\usepackage{booktabs}
\usepackage{listings}
\usepackage{fancyvrb}
\usepackage{microtype}
\usepackage{hyperref}

\DeclareGraphicsExtensions{.pdf,.png,.jpg,.eps}

\newcommand{\be}{\begin{equation}}
\newcommand{\ee}{\end{equation}}
\newcommand{\cpk}{CP2K/Quickstep}
\newcommand{\cpdg}{second-generation Car--Parrinello}
\renewcommand{\doi}[1]{\href{https://doi.org/#1}{doi:#1}}
\newcommand{\githubrepoURL}{https://github.com/DCM-Uni-Paderborn/CP2G-Decalogo}
\newcommand{\githubrepository}{\href{\githubrepoURL}{\texttt{DCM-Uni-Paderborn/CP2G-Decalogo}}}
\newcommand{\cpincludegraphics}[2][]{%
  \IfFileExists{figures/#2.pdf}{\includegraphics[#1]{figures/#2.pdf}}{%
    \IfFileExists{#2-eps-converted-to.pdf}{\includegraphics[#1]{#2-eps-converted-to.pdf}}{\includegraphics[#1]{#2.eps}}%
  }%
}

\hypersetup{
  pdftitle={Best practices for second-generation Car-Parrinello ab initio molecular dynamics with CP2K/Quickstep},
  pdfauthor={Thomas D. Kuehne},
  pdfkeywords={CP2K, Quickstep, second-generation Car-Parrinello, ab initio molecular dynamics, modified Langevin dynamics}
}

\begin{document}

\title{Best practices for \cpdg{} \textit{ab initio} molecular dynamics with \cpk{}}

\author{Thomas D. K\"uhne}
\email{tkuehne@cp2k.org}
\affiliation{Center for Advanced Systems Understanding (CASUS), Conrad-Schiedt-Stra{\ss}e 20, 02826 G\"orlitz, Germany}
\affiliation{Helmholtz Zentrum Dresden-Rossendorf, Bautzner Landstra{\ss}e 400, 01328 Dresden, Germany}
\affiliation{Institute of Artificial Intelligence, Technische Universit\"at Dresden, Helmholtzstra{\ss}e 10, 01069 Dresden, Germany}

\date{May 25, 2026}

\begin{abstract}

Second-generation Car--Parrinello \textit{ab initio} molecular dynamics (CP2G AIMD) combines a Born--Oppenheimer-like nuclear equation of motion with a predictor-corrector propagation of the one-particle density matrix and a modified Langevin equation to ensure an accurate sampling of the Boltzmann distribution.  In the \cpk{} implementation, this approach can substantially reduce the number of self-consistent-field iterations required per molecular-dynamics step, but only if the density-matrix propagation, the orbital-transformation corrector, and the modified Langevin friction parameters are tuned consistently.  This tutorial gives a practical workflow for setting up CP2G AIMD calculations in \cpk{}, starting from Born--Oppenheimer pre-equilibration and proceeding through corrector-step-size selection, always-stable predictor-corrector order optimization, final relaxation, and the calibration of the dissipative and stochastic terms used for canonical sampling.  The protocol is demonstrated for a 32-molecule liquid-water benchmark and is written for users who already know the basics of \cpk{} input files but want a reliable route from a conventional AIMD setup to a production-ready CP2G simulation.
\end{abstract}

\maketitle

\section{Scope and intended use}

This article is a practical guide to running second-generation Car--Parrinello (CP2G) \textit{ab initio} molecular dynamics (AIMD) in \cpk{}.  It does not replace the formal derivations of the CP2G method, the orbital-transformation minimizer, or the modified Langevin correction; instead, it collects the operational choices that most often determine whether a calculation is stable, efficient, and statistically meaningful.

The maintained article source, tutorial inputs, and future updates are hosted in the public GitHub repository \githubrepository.  Readers are encouraged to report ambiguities, version-dependent behavior, or suggested extensions through the issue tracker of that repository.

\section{Prerequisites and software requirements}

This tutorial assumes familiarity with standard MD concepts, basic DFT terminology, and the structure of CP2K input files.  Users should be comfortable running short \cpk{} calculations from the command line and inspecting energies, temperatures, and restart files.  No programming is required for the workflow itself, although plotting the diagnostic trajectories is useful.

The input syntax follows CP2K version 2026.1 and the corresponding online manual.  Users who run the tutorial with a later CP2K release should record the CP2K version or git commit, the compiler/MPI environment, and any changes to keyword names or default values.  The examples require a \cpk{} executable with the Quickstep module, Gaussian-and-plane-wave DFT, the orbital-transformation minimizer, and ASPC density-matrix extrapolation enabled.

\section*{Learning objectives}
\begin{itemize}
\item Prepare a conventional Born--Oppenheimer molecular-dynamics pre-equilibration that also writes the density-matrix history required for clean CP2G restarts.
\item Determine a stable orbital-transformation corrector step size and an appropriate always-stable predictor-corrector order.
\item Relax the system with the final CP2G density-matrix-propagation settings before production sampling.
\item Tune the dissipative and stochastic parameters of the modified Langevin equation so that residual non-self-consistent forces do not bias canonical averages.
\item Diagnose common failure modes from total-energy drift, orbital-transformation convergence, and species-resolved temperatures.
\end{itemize}

\section{Introduction}

Molecular dynamics (MD) simulations \cite{AlderWainwright1957, Rahman1964} provide static and dynamic equilibrium properties from finite-temperature trajectories.  For density-functional-theory (DFT)-based \textit{ab initio} molecular dynamics (AIMD), the forces are obtained on the fly from the electronic structure, so the cost per time step is usually dominated by the self-consistent-field (SCF) optimization.  The CP2G method \cite{TDK2007, TDK2014, Prodan2018} reduces this cost by propagating the one-particle density matrix (DM) and using it to build an approximate Kohn--Sham (KS) Hamiltonian \cite{HohenbergKohn1964, KohnSham1965}.  Because the propagated DM is not, in general, identical to the fully self-consistent ground-state DM, the final equilibration and production run must use the same density-matrix-propagation (DMP) parameters.

The formal background is given in the original CP2G paper \cite{TDK2007}, the WIREs review \cite{TDK2014}, and the discussion of configuration-space sampling in disordered crystals \cite{Prodan2018}.  Independent analyses and applications support the methodological basis: Hutter placed Car--Parrinello and Born--Oppenheimer dynamics in a common framework, Dai and Yuan rationalized the modified Langevin correction for noisy forces, Scheiber, Shi, and Khaliullin used the same idea to stabilize absolutely localized molecular-orbital (ALMO) AIMD in CP2K, and Musso et al. demonstrated CP2G for the hexagonal boron nitride/Rh(111) (h-BN/Rh(111)) nanomesh with a reported 17-fold speed-up \cite{Hutter2012,DaiYuan2009,Scheiber2018,Musso2018}.  For users of \cpk{}, the \cpk{} theory and code paper \cite{CP2K}, the 2026 practical overview of the CP2K program package \cite{CP2KSimple2026}, and the AIMD guide by Hutter, Iannuzzi, and K\"uhne \cite{HutterIannuzziKuehne2024} are strongly recommended companion references.  The latter two are particularly useful because they place CP2G in the broader context of practical CP2K usage and modern AIMD workflows.

The most robust practical route begins with a short DFT-level Born--Oppenheimer MD (BOMD) pre-equilibration \cite{TDK2014, Payne1992}.  A massive Nos\'e--Hoover thermostat \cite{Nose1984, Hoover1985}, a high-friction Langevin thermostat using the Ricci--Ciccotti integrator \cite{RicciCiccotti2003}, or the stochastic velocity-rescaling thermostat of Bussi, Donadio, and Parrinello \cite{Bussi2007} are all suitable choices.  For the Ricci--Ciccotti integrator, values around $1/20 \gtrsim \gamma \Delta t \gg 1/1000$ provide a useful high-friction pre-equilibration regime.

To demonstrate the workflow, we use the standard 32-molecule liquid-water benchmark at ambient conditions ($T=300$~K, $P=1$~bar, $a=9.8528$~\AA) \cite{TDKwater2009}.  The starting geometry is taken from \texttt{/cp2k/tests/QS/benchmark/H2O-32.inp}, which was equilibrated with the TIP5P force field \cite{MahoneyJorgensen2000}.  The Goedecker--Teter--Hutter (GTH) pseudopotentials \cite{GTH1996, HGH1998, Krack2005} and the corresponding Gaussian basis sets are reproduced in the Appendix.  The numerical setup is intentionally modest so that the tutorial remains lightweight; production calculations should repeat the same workflow with basis sets, plane-wave cutoffs, and simulation lengths appropriate to the target observable.

\section{Born--Oppenheimer pre-equilibration}
\label{part0}

The initial BOMD reads as follows (please consult the online manual at \texttt{https://manual.cp2k.org} for details). 
\begin{verbatim}
&FORCE_EVAL
  METHOD QS
  &DFT
    BASIS_SET_FILE_NAME ./H2O.qbs
    POTENTIAL_FILE_NAME ./GTH_POTENTIALS
    &MGRID
      CUTOFF 240
    &END MGRID
    &QS
      EPS_DEFAULT 1.0E-12
      EXTRAPOLATION ASPC
      EXTRAPOLATION_ORDER 3
      MAP_CONSISTENT TRUE
    &END QS
    &SCF
      EPS_SCF 5.0E-7
      MAX_SCF 25
      &OT ON
        MINIMIZER DIIS
        STEPSIZE 0.15
      &END OT
      &OUTER_SCF
        EPS_SCF 5.0E-7
        MAX_SCF 40
      &END OUTER_SCF
      SCF_GUESS RESTART
     #MAX_SCF_HISTORY 1
      &PRINT
        &RESTART_HISTORY
          EACH 1 1 0
          FILENAME =RESTART
          BACKUP_COPIES 5
        &END RESTART_HISTORY
      &END PRINT
    &END SCF
    &XC
      &XC_FUNCTIONAL PBE
      &END XC_FUNCTIONAL
    &END XC
  &END DFT
  &SUBSYS
    &CELL
      ABC 9.8528 9.8528 9.8528
      UNIT ANGSTROM
    &END CELL
    # 32 H2O (TIP5P,1bar,300K) a = 9.8528
    &COORD
    &END COORD
    &KIND H
      BASIS_SET DZVP-GTH
      POTENTIAL GTH-PBE-q1
    &END KIND
    &KIND O
      BASIS_SET DZVP-GTH
      POTENTIAL GTH-PBE-q6
    &END KIND
    &TOPOLOGY
      CONNECTIVITY OFF
    &END TOPOLOGY
  &END SUBSYS
&END FORCE_EVAL
&GLOBAL
  PROJECT H2O-32
  RUN_TYPE MD
  PRINT_LEVEL LOW
  WALLTIME 28000
&END GLOBAL
&MOTION
  &MD
    ENSEMBLE NVT
    STEPS 10000
    TIMESTEP 0.5
    TEMPERATURE 300.0
    &THERMOSTAT
      TYPE NOSE
      REGION MASSIVE
      &NOSE
        LENGTH 3
        YOSHIDA 3
        TIMECON 100.0
        MTS 2
      &END NOSE
    &END THERMOSTAT
  &END MD
&END MOTION
&EXT_RESTART
  RESTART_FILE_NAME H2O-32-1.restart
&END EXT_RESTART
\end{verbatim}

The \texttt{RESTART\_HISTORY} block is essential for a clean transition from BOMD to CP2G.  It must store enough previous DMs for the chosen always-stable predictor-corrector (ASPC) order.  With \texttt{EXTRAPOLATION ASPC}, \texttt{BACKUP\_COPIES} should therefore be set to \texttt{EXTRAPOLATION\_ORDER + 2}.  The setting \texttt{EACH 1 1 0} writes DMs only after completed AIMD steps and not during the intermediate SCF cycles, which avoids inconsistent history restarts.
\begin{figure*}[t]
  \cpincludegraphics[width=\linewidth]{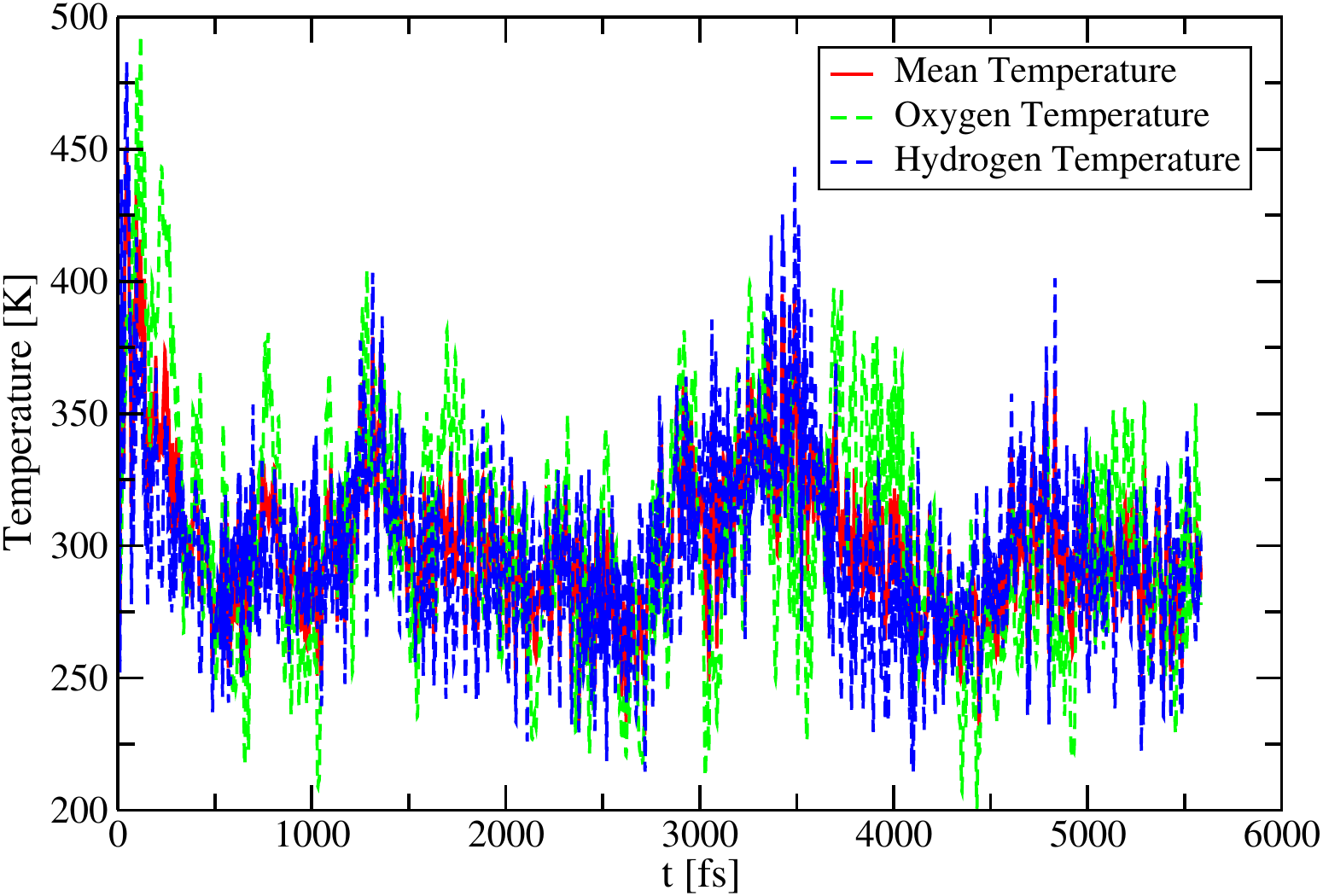}
  \caption{BOMD pre-equilibration of the 32-molecule water benchmark.  The plot shows the instantaneous nuclear temperature during the initial NVT run; at this stage the trajectory only needs to be sufficiently stable and equilibrated to seed the subsequent CP2G parameter search.}
  \label{fig:bomd}
\end{figure*}
At this stage, the nuclear temperature should show the behavior illustrated in Figure~\ref{fig:bomd}.  If the initial structure is already equilibrated at the DFT level and a consistent DM history is available, this pre-equilibration step can be shortened or omitted.

\section{Optimizing the density-matrix propagation}
\label{part1}

The CP2G setup contains two conceptually separate optimizations.  First, the DMP parameters are chosen so that the propagated electronic structure remains close enough to the Born--Oppenheimer (BO) surface for the target accuracy.  Second, the modified Langevin equation is tuned so that the remaining dissipative error does not bias canonical sampling.  The first step controls the quality of the potential-energy surface and, more importantly, the nuclear forces; the second step controls the sampled Boltzmann distribution.

The DMP optimization itself has two small, largely independent tasks: selecting the corrector step size and identifying the optimal order $K$ of the ASPC integrator \cite{Kolafa2004, Kolafa2005}.  In \cpk{}, \texttt{EXTRAPOLATION\_ORDER k} corresponds to $K+2$.  In practice, this reduces the search to a small set of short trial calculations rather than a high-dimensional parameter scan.

\subsection{Choosing the corrector step size}

At first sight this step may appear redundant, because the corrector step size resembles the parameter $\omega$ in Ref.~[\onlinecite{TDK2007}].  In \cpk{}, however, the relevant value is affected by the orbital-transformation (OT) minimizer and by its preconditioner.

The parameter $\omega$ is defined on the Grassmann manifold of idempotent DMs \cite{EdelmanArias1998}, or equivalently on the manifold of orthonormal KS orbitals.  In the original Car--Parrinello molecular-dynamics (CPMD) approach \cite{CarParrinello1985}, this idempotency constraint is enforced explicitly at every step.  In \cpk{}, the OT method \cite{OT2003}, inspired by the exponential transformation of Hutter, Parrinello, and Vogel \cite{HutterParrinello1994}, parameterizes the orbitals through an auxiliary variable in a tangent space.  Idempotency is therefore preserved even when the electronic minimization is incomplete, while the preconditioner \cite{Gan2000} determines how efficiently the auxiliary variables approach the minimum.

The practical consequence is that the value used in the input, \texttt{STEPSIZE}, corresponds to an implementation-specific $\omega'$ rather than to the bare theoretical $\omega$.  The default \texttt{STEPSIZE} is 0.15, but stable values often lie between 0.05 and 0.15.  The goal is not to overfit this parameter; it is to find a value that keeps the propagation stable and minimizes the total-energy drift for the chosen number of corrector steps.

Run a small set of short NVE exploration trajectories, typically 3--7 runs of a few hundred steps, each restarted from the BOMD relaxation.  Keep all settings fixed except \texttt{STEPSIZE}, and check both the energy drift and the OT convergence criterion.  The \texttt{RESTART\_HISTORY} written during BOMD is used here by setting \texttt{SCF\_GUESS HISTORY\_RESTART}; this reads the stored DMs and predicts the first DM needed for the first force in the velocity-Verlet loop.  If \texttt{MAX\_SCF\_HISTORY} is larger than one, the previous MD step must have completed cleanly before the restart.

For this scan, activate CP2G with \texttt{EXTRAPOLATION ASPC} and choose a low \texttt{EXTRAPOLATION\_ORDER}, for example 0 or 1.  This deliberately increases the energy loss, but it makes differences between \texttt{STEPSIZE} values easier to detect.  The number of corrector steps is controlled by \texttt{MAX\_SCF\_HISTORY}; values of 1 or 2 are often sufficient, but the required value is system- and accuracy-dependent.

Finally, switch the trajectory to the microcanonical (NVE) ensemble so that the total-energy drift can be monitored without thermostat feedback.  The relevant input changes are:

\begin{verbatim}
&DFT
  RESTART_FILE_NAME RESTART
  &QS
    EXTRAPOLATION ASPC
    EXTRAPOLATION_ORDER 3
  &END QS
  &SCF
    &OT ON
      MINIMIZER DIIS
      STEPSIZE 0.125
    &END OT
    MAX_SCF_HISTORY 1
    SCF_GUESS HISTORY_RESTART
    &PRINT
      &RESTART_HISTORY
        EACH 1 1 0
        FILENAME =RESTART
        BACKUP_COPIES 5
      &END RESTART_HISTORY
    &END PRINT
  &END SCF
  &MD
    ENSEMBLE NVE
  &END MD
&END DFT
\end{verbatim}

\begin{figure*}[t]
  \cpincludegraphics[width=\linewidth]{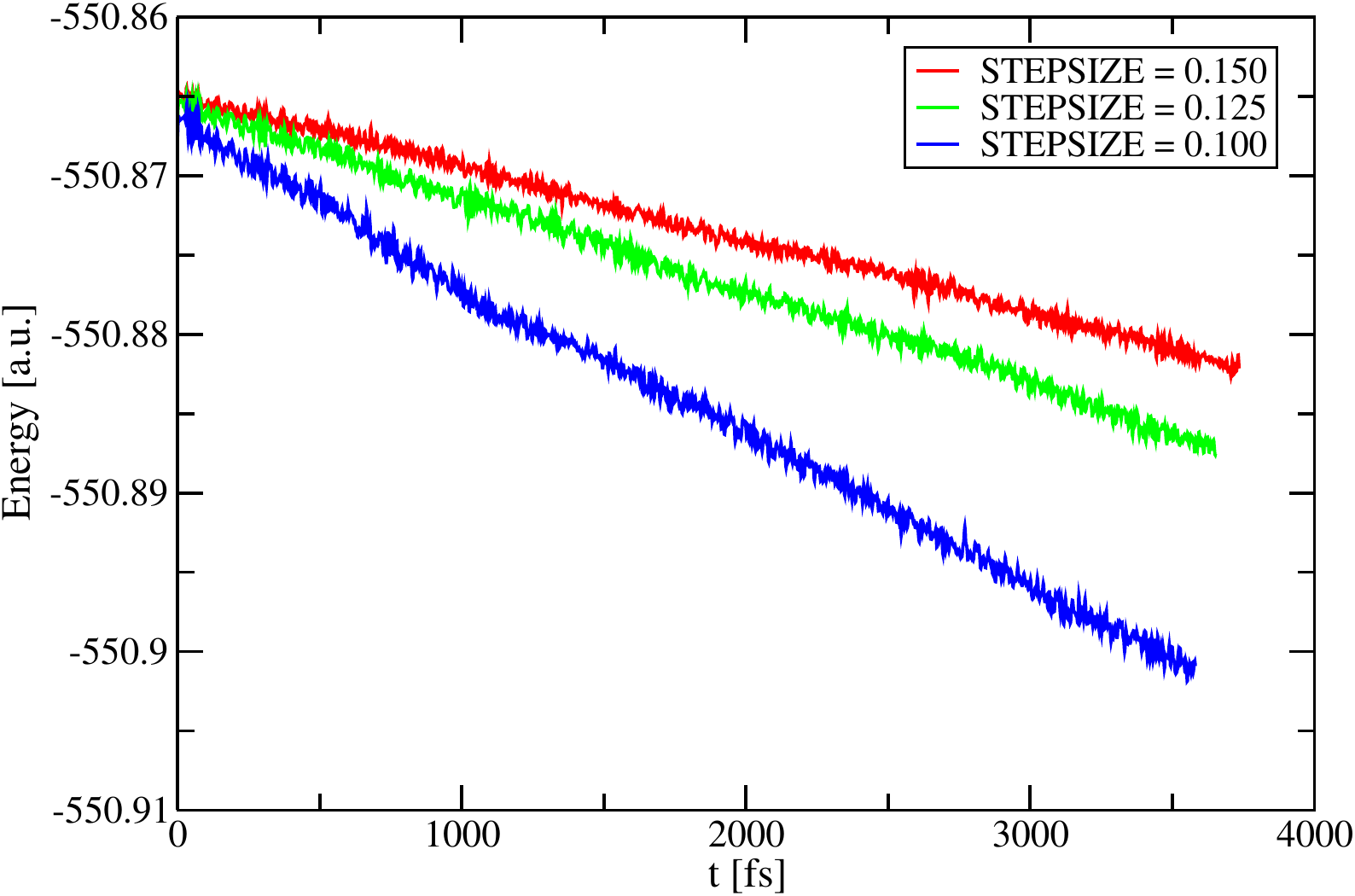}
  \caption{Selection of the OT corrector step size.  The optimal \texttt{STEPSIZE} is identified from short NVE trajectories by minimizing the total-energy drift while retaining stable OT convergence.}
  \label{fig:ot-stepsize}
\end{figure*}

For the present water benchmark, Figure~\ref{fig:ot-stepsize} supports \texttt{STEPSIZE 0.15}, which is therefore kept fixed in the following scans.

\subsection{Choosing the ASPC propagation order}

The accuracy of the propagated electronic structure depends on the short-time integration of the electronic degrees of freedom.  This is distinct from the nuclear trajectory, where long-time accuracy is limited by the Lyapunov instability of MD.  CP2G therefore uses an explicit Gear-type ASPC integrator \cite{Gear1971} for the electronic degrees of freedom rather than integrating nuclei and electrons on exactly the same footing as in CPMD.

ASPC is not fully time-reversible; for order $K$, the time-reversibility error is pushed to high order in the MD time step $h$.  Time reversibility is desirable because the underlying Hamiltonian dynamics is time-reversible, but the more consequential issue here is non-symplecticity, which manifests as a systematic energy drift.

The optimal ASPC order is found in the same way as the corrector step size: restart several short NVE trajectories, keep the previously selected \texttt{STEPSIZE}, vary only \texttt{EXTRAPOLATION\_ORDER k}, and choose the smallest-drift setting that does not degrade the OT convergence.  Values of \texttt{k} between 0 and 3 are typical.  More disordered systems, smaller band gaps, higher temperatures, and larger nuclear time steps often favor lower orders.

\begin{figure*}[t]
  \cpincludegraphics[width=\linewidth]{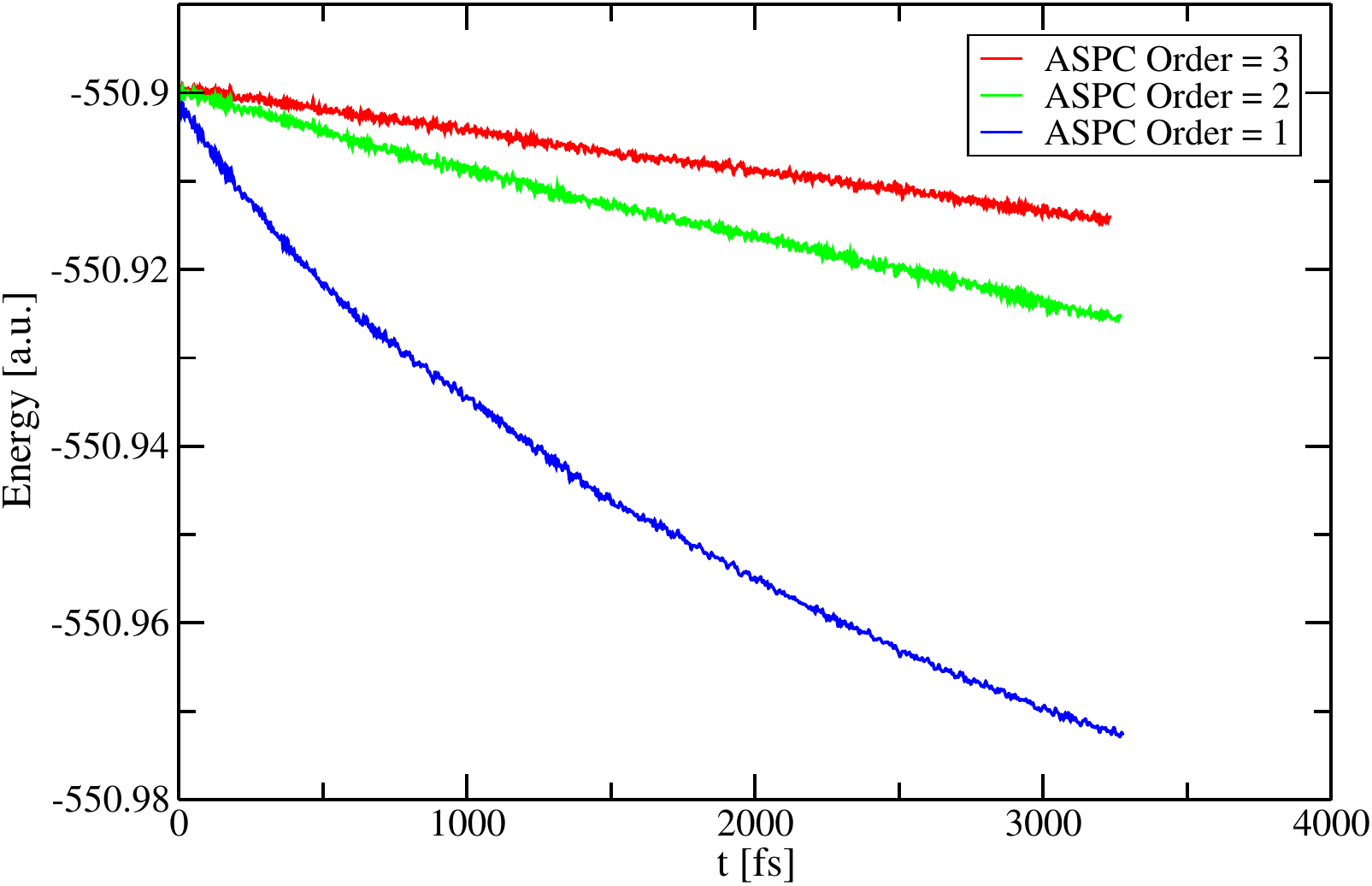}
  \caption{Effect of the ASPC propagation order on the total-energy evolution.  The preferred order is the one that gives the smallest systematic drift without compromising the electronic corrector convergence.}
  \label{fig:aspc-order}
\end{figure*}

For the present benchmark, Figure~\ref{fig:aspc-order} shows that the initial choice \texttt{EXTRAPOLATION\_ORDER 3} is already optimal.  This value is fixed for all subsequent relaxation and sampling runs.

\section{Final CP2G relaxation}

After the preceding scans, the DMP accuracy and therefore the deviation from the BO surface are fixed.  The energy drift has been minimized for the chosen number of corrector steps, but the resulting trajectory is not yet a production trajectory.  The system should first be relaxed with exactly the DMP parameters that will be used later for sampling.  This final relaxation is the step that makes the propagated DM history, nuclear coordinates, and velocities mutually consistent.

The relevant quantity is not the absolute value of the electronic energy functional, but the residual non-self-consistent force \cite{MarxHutter2000}.  In CPMD, the electronic equation of motion, inspired by Ehrenfest dynamics \cite{Ehrenfest1927}, is constructed so that the instantaneous electronic state yields consistent forces.  CP2G instead uses a BO-like nuclear equation of motion without a fictitious electronic mass.  Residual force errors must therefore be reduced by the corrector and the remaining dissipative component must be compensated statistically by the modified Langevin equation.

The Hamiltonian matrix is now built from the propagated density.  If the parameter search was successful, it is close to the self-consistent KS Hamiltonian, but it is still not identical to it.  The final relaxation is therefore performed with the fast CP2G settings by switching to \texttt{ENSEMBLE LANGEVIN}:

\begin{verbatim}
  &MD
    ENSEMBLE LANGEVIN
    GAMMA 0.01
  &END MD
\end{verbatim}

\begin{figure*}[t]
  \cpincludegraphics[width=\linewidth]{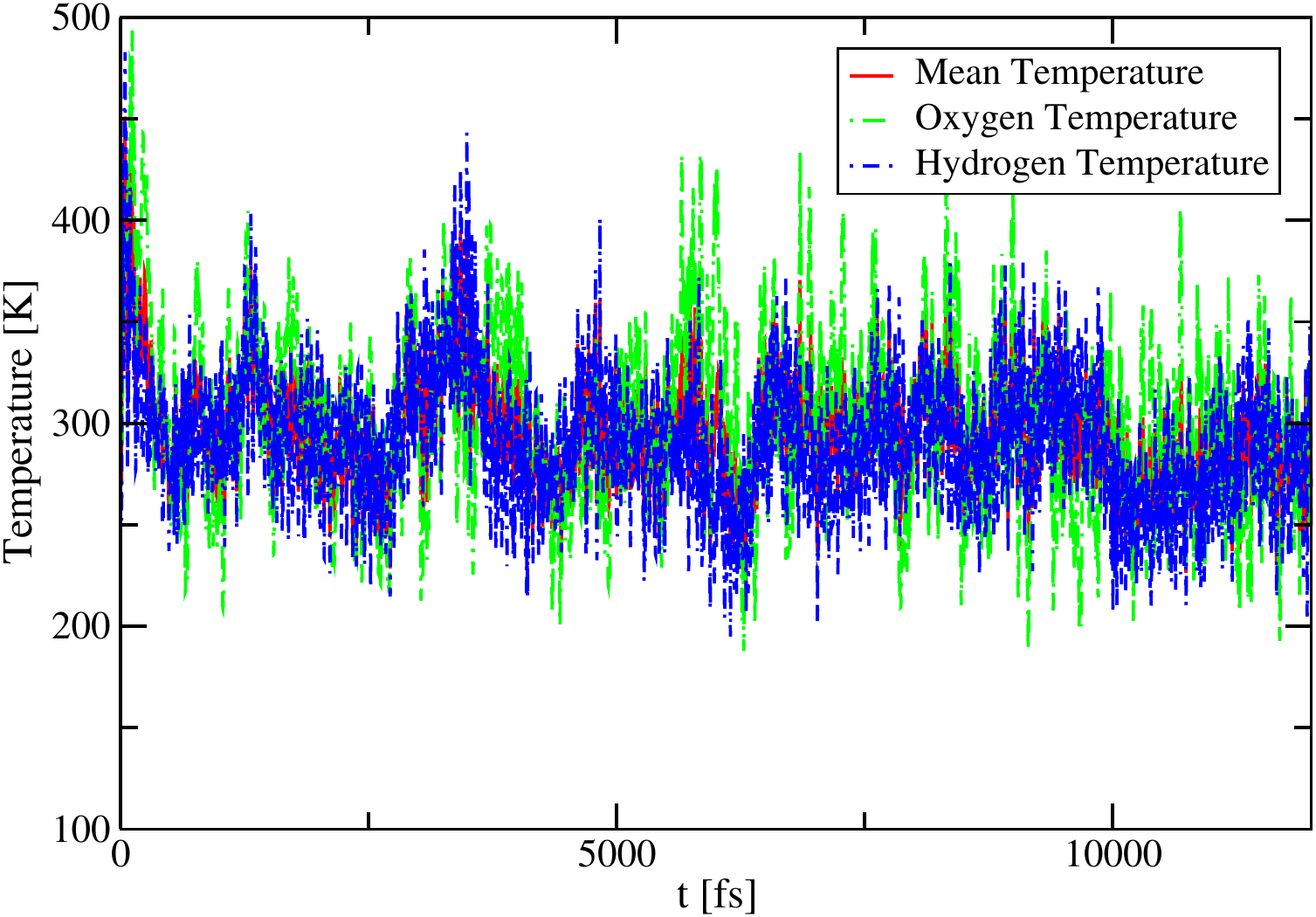}
  \caption{Final relaxation with the selected CP2G settings.  The trajectory begins from the preceding BOMD relaxation and then continues with the fast CP2G propagation so that the propagated density, coordinates, and velocities relax together.}
  \label{fig:final-relaxation}
\end{figure*}

In Figure~\ref{fig:final-relaxation}, only the first 2.8~ps correspond to the original BOMD relaxation; the remainder uses the CP2G settings described above.  The species-resolved temperatures remain equilibrated, and the different amplitudes mainly reflect the stoichiometry of the system.

\section{Tuning the modified Langevin equation}
\label{part2}

The remaining task is to ensure correct sampling.  A non-symplectic predictor-corrector propagation introduces a dissipative component, visible as total-energy drift in NVE tests.  In CP2G this dissipative error is compensated by a modified Langevin equation.  Although the mean temperature is the primary tuning target, species-resolved temperatures provide a more sensitive diagnostic and should be checked whenever possible.

The method therefore samples the canonical rather than the microcanonical ensemble.  This is usually not a practical limitation as long as the total friction $\gamma=\gamma_D+\gamma_L$ remains small compared with the inverse relaxation time of the system.  The dissipative contribution $\gamma_D$ is system-dependent and reflects the residual CP2G error at the chosen number of corrector steps.  Increasing \texttt{MAX\_SCF\_HISTORY} reduces $\gamma_D$ and continuously approaches BOMD.  In that sense, the number of corrector steps plays a role analogous to the fictitious electronic mass in CPMD, where both the maximum time step \cite{PastoreBuda1991} and the deviation from the BO surface \cite{BornemannSchuette1998} scale with $\sqrt{\mu}$.

For canonical averages, it is usually sufficient to determine \texttt{NOISY\_GAMMA} ($\gamma_D$) to within a useful range and to add a modest overlay friction \texttt{GAMMA} ($\gamma_L$).  A painstakingly precise determination of \texttt{NOISY\_GAMMA} is only needed when the trajectory must remain as close as possible to microcanonical dynamics.

\subsection[Bootstrapping gammaD]{Bootstrapping $\gamma_D$}

Start with a coarse scan to identify the order of magnitude of \texttt{NOISY\_GAMMA}.  A representative input fragment is:

\begin{verbatim}
  &MD
    ENSEMBLE LANGEVIN
    GAMMA 0.0
    NOISY_GAMMA 0.0001
  &END MD
\end{verbatim}

The OT convergence criterion from the final CP2G relaxation provides a useful starting estimate: lower preconditioned mean-gradient deviations usually correspond to smaller \texttt{NOISY\_GAMMA}.  Here, five trial values are distributed over the initial range $10^{-3}$--$10^{-5}$~fs$^{-1}$.

\begin{figure*}[t]
  \cpincludegraphics[width=\linewidth]{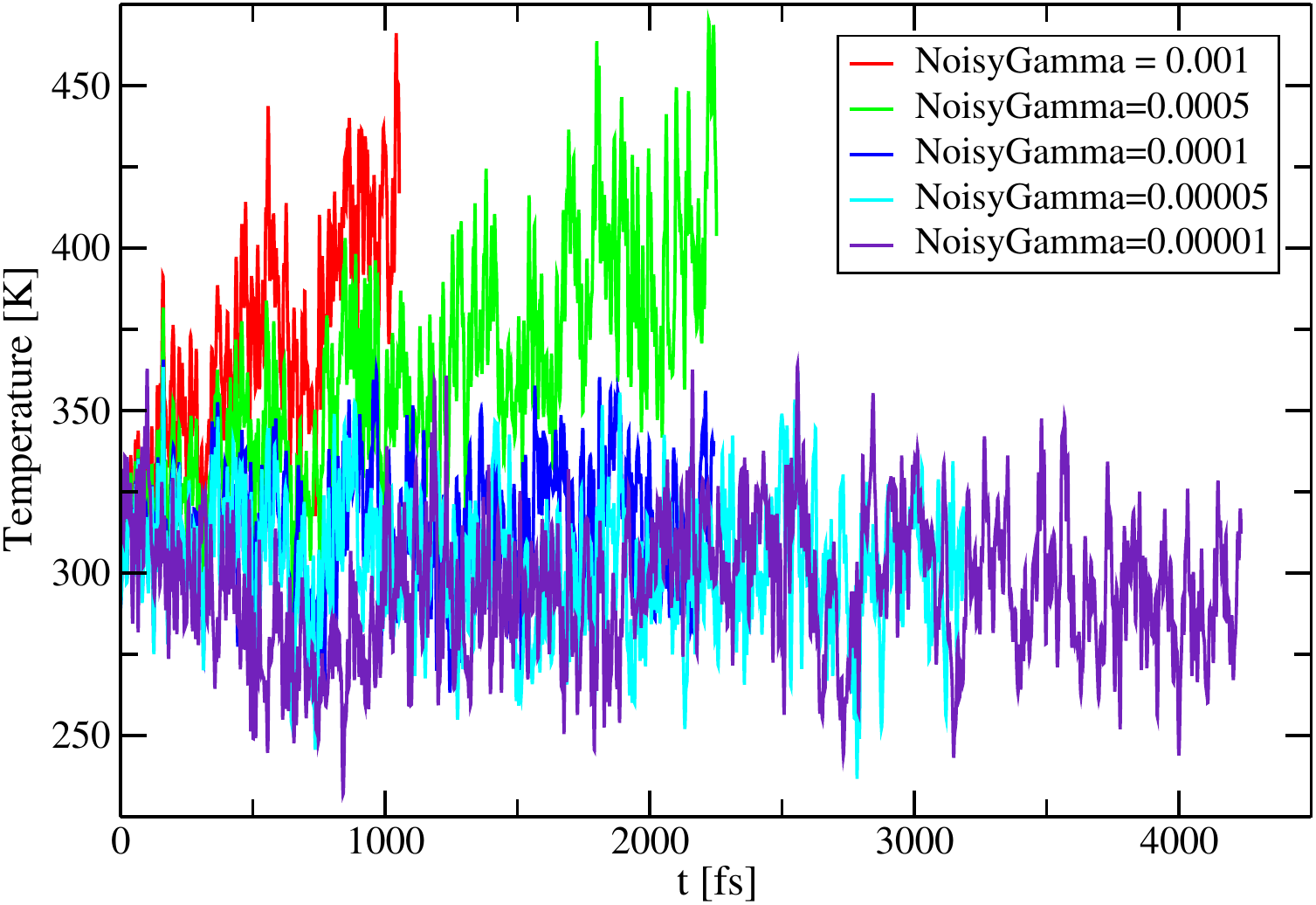}
  \caption{Coarse scan of the dissipative Langevin parameter $\gamma_D$.  Trial values spanning $10^{-3}$--$10^{-5}$~fs$^{-1}$ identify the range in which the mean temperature remains closest to the target value.}
  \label{fig:coarse-gamma}
\end{figure*}

\begin{figure*}[t]
  \cpincludegraphics[width=\linewidth]{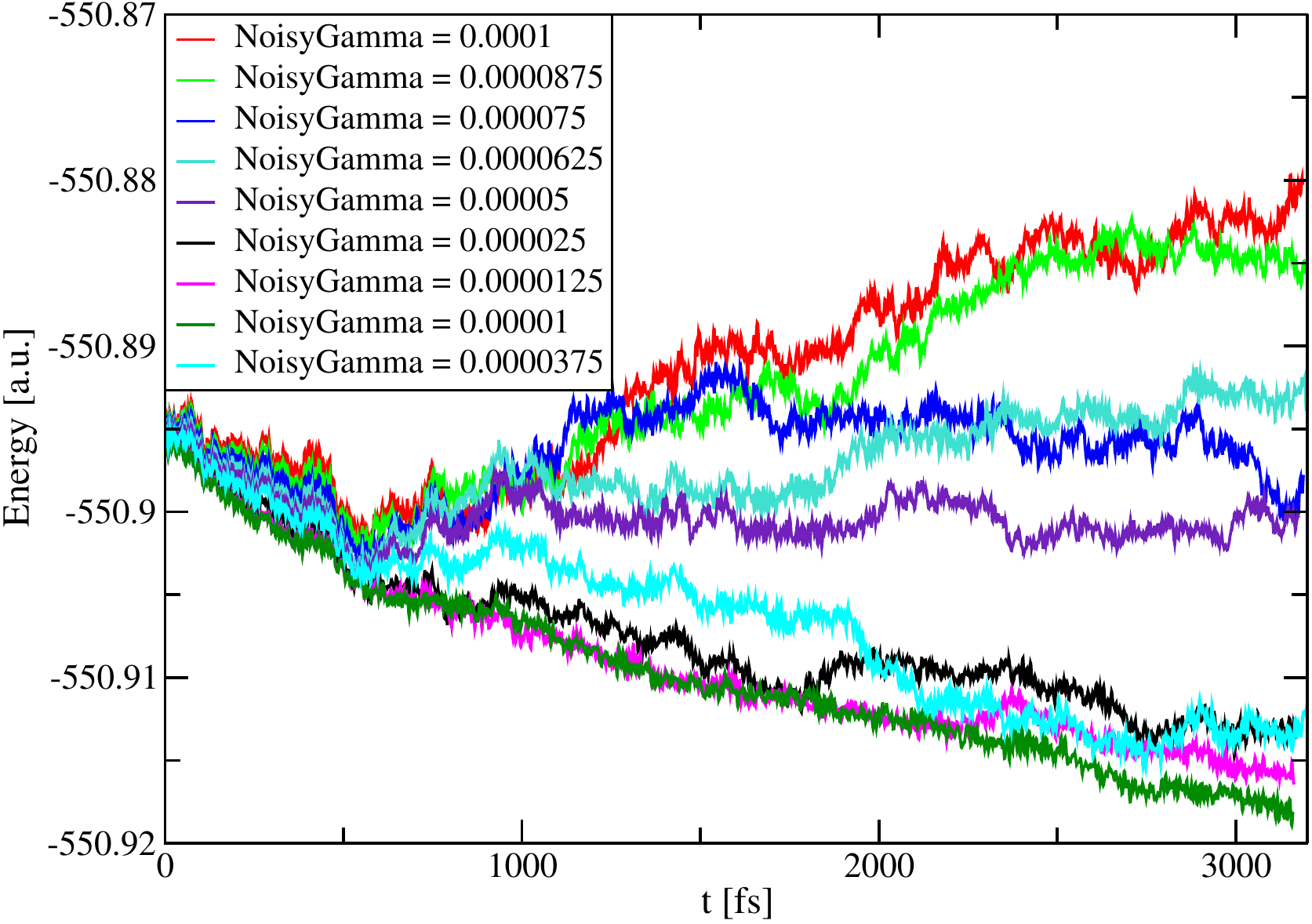}
  \caption{Total-energy evolution in the fine scan of $\gamma_D$.  The curves are used together with the temperature traces in Figure~\ref{fig:fine-gamma-temperature} to select a dissipative correction that compensates the CP2G drift without over-thermostatting the dynamics.}
  \label{fig:fine-gamma-energy}
\end{figure*}

\begin{figure*}[t]
  \cpincludegraphics[width=\linewidth]{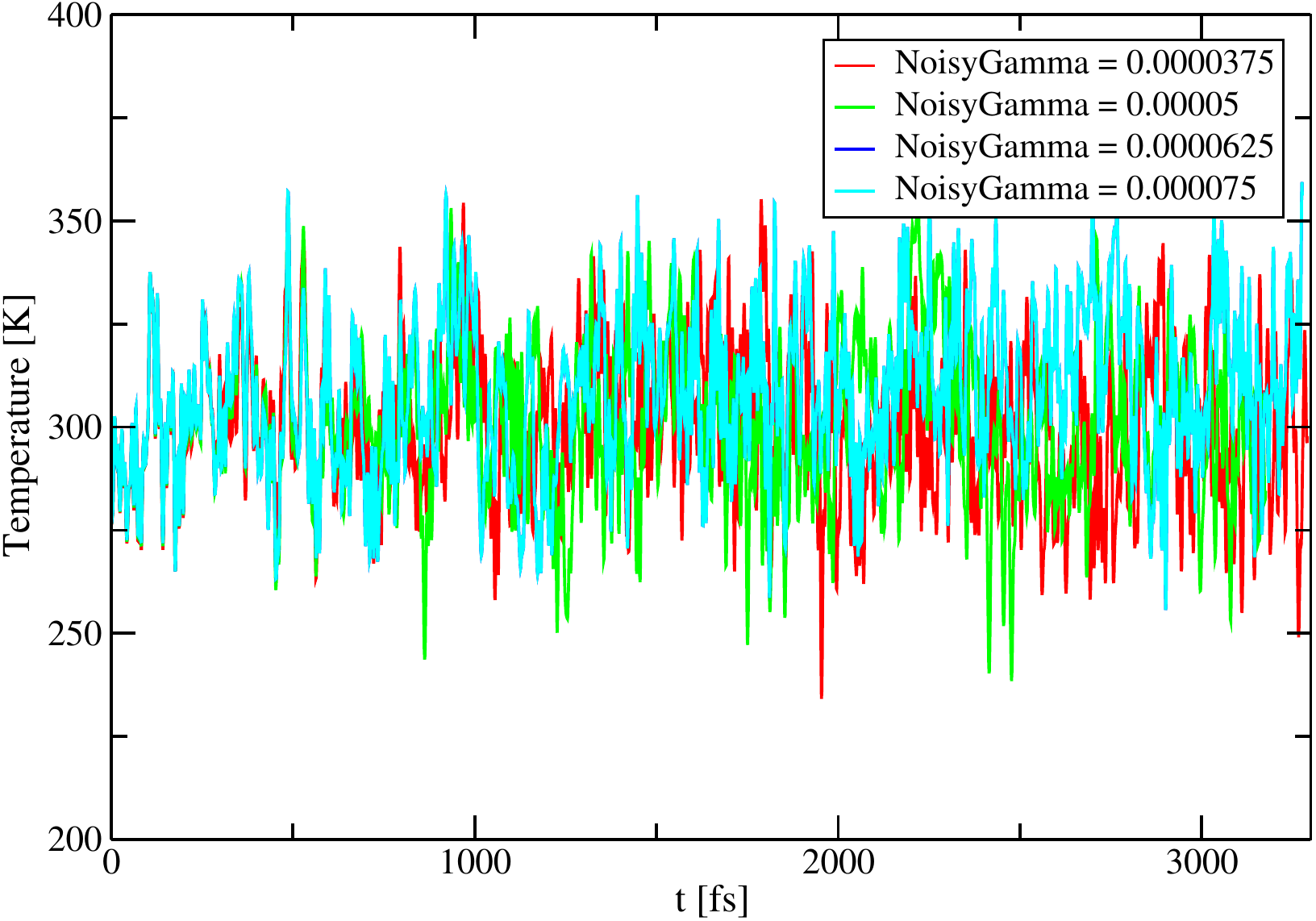}
  \caption{Temperature evolution in the fine scan of $\gamma_D$.  The best value keeps the mean temperature close to the target while avoiding systematic heating or cooling over the trajectory.}
  \label{fig:fine-gamma-temperature}
\end{figure*}

Figure~\ref{fig:coarse-gamma} suggests that the fine scan can be restricted to $10^{-4}$--$10^{-5}$~fs$^{-1}$.  We therefore restart nine additional trajectories from the continued CP2G relaxation and distribute them evenly across this narrower interval (Figures~\ref{fig:fine-gamma-energy} and \ref{fig:fine-gamma-temperature}).

\begin{figure*}[t]
  \cpincludegraphics[width=\linewidth]{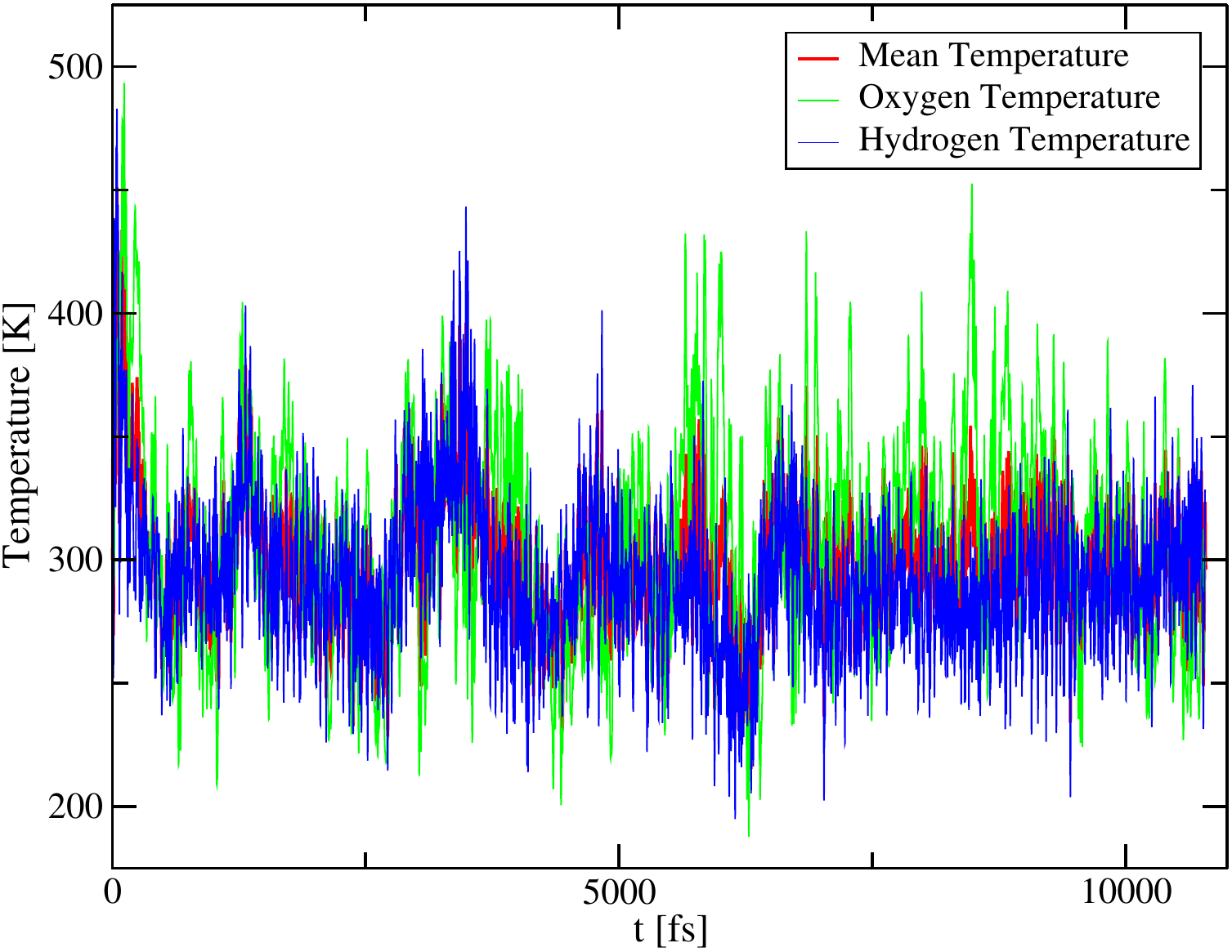}
  \caption{Species-resolved temperatures using $\gamma_D=5\times10^{-5}$~fs$^{-1}$.  Agreement between the hydrogen and oxygen temperatures is a stringent test that the dissipative correction is not biasing the canonical distribution.}
  \label{fig:species-temperature}
\end{figure*}

For this benchmark, the fine scan supports \texttt{NOISY\_GAMMA 0.00005}.  The choice is then checked by monitoring the species-resolved temperatures (Figure~\ref{fig:species-temperature}).

The agreement is already satisfactory.  The remaining mismatch reflects the finite precision of the bootstrapped $\gamma_D$, so a nonzero overlay friction $\gamma_L$ is recommended in production runs.

\subsection{Adding overlay noise for canonical sampling}

Here, two overlay frictions are tested: $\gamma_L=10^{-4}$~fs$^{-1}$ and $\gamma_L=3.75\times10^{-4}$~fs$^{-1}$.  Even the lower value equilibrates the atomic species for this benchmark, but production calculations should not use an overly small $\gamma_L$ unless the species-resolved temperatures have been checked explicitly.  The appropriate value is system-dependent; in practice, values above $10^{-4}$~fs$^{-1}$ can often be used without a noticeable effect on the relevant dynamics.

\begin{figure*}[t]
  \cpincludegraphics[width=\linewidth]{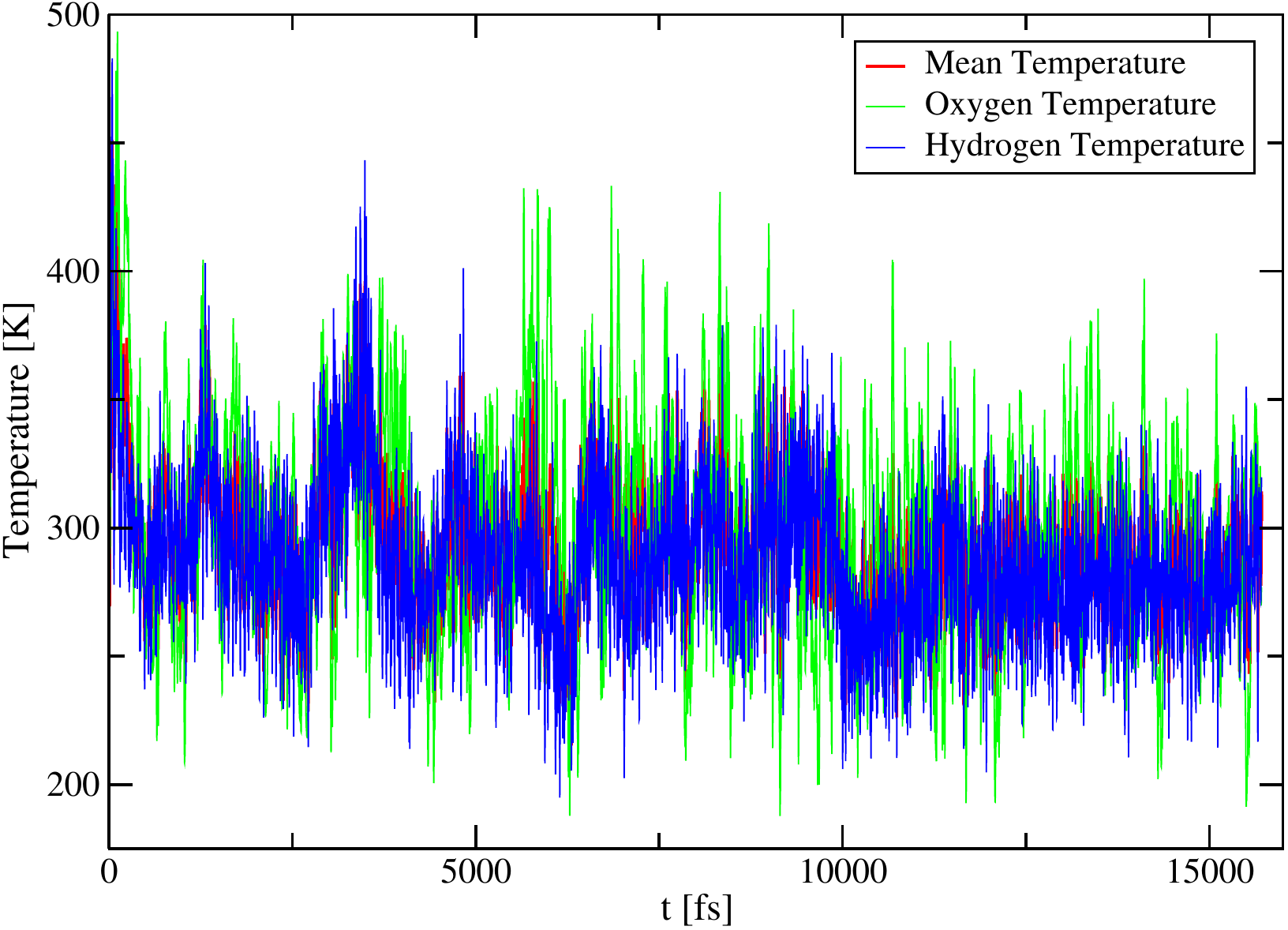}
  \caption{Species-resolved temperature evolution with overlay friction $\gamma_L=10^{-4}$~fs$^{-1}$.  The overlay noise removes the remaining imbalance after the dissipative correction has been bootstrapped.}
  \label{fig:gamma-l-1000}
\end{figure*}

\begin{figure*}[t]
  \cpincludegraphics[width=\linewidth]{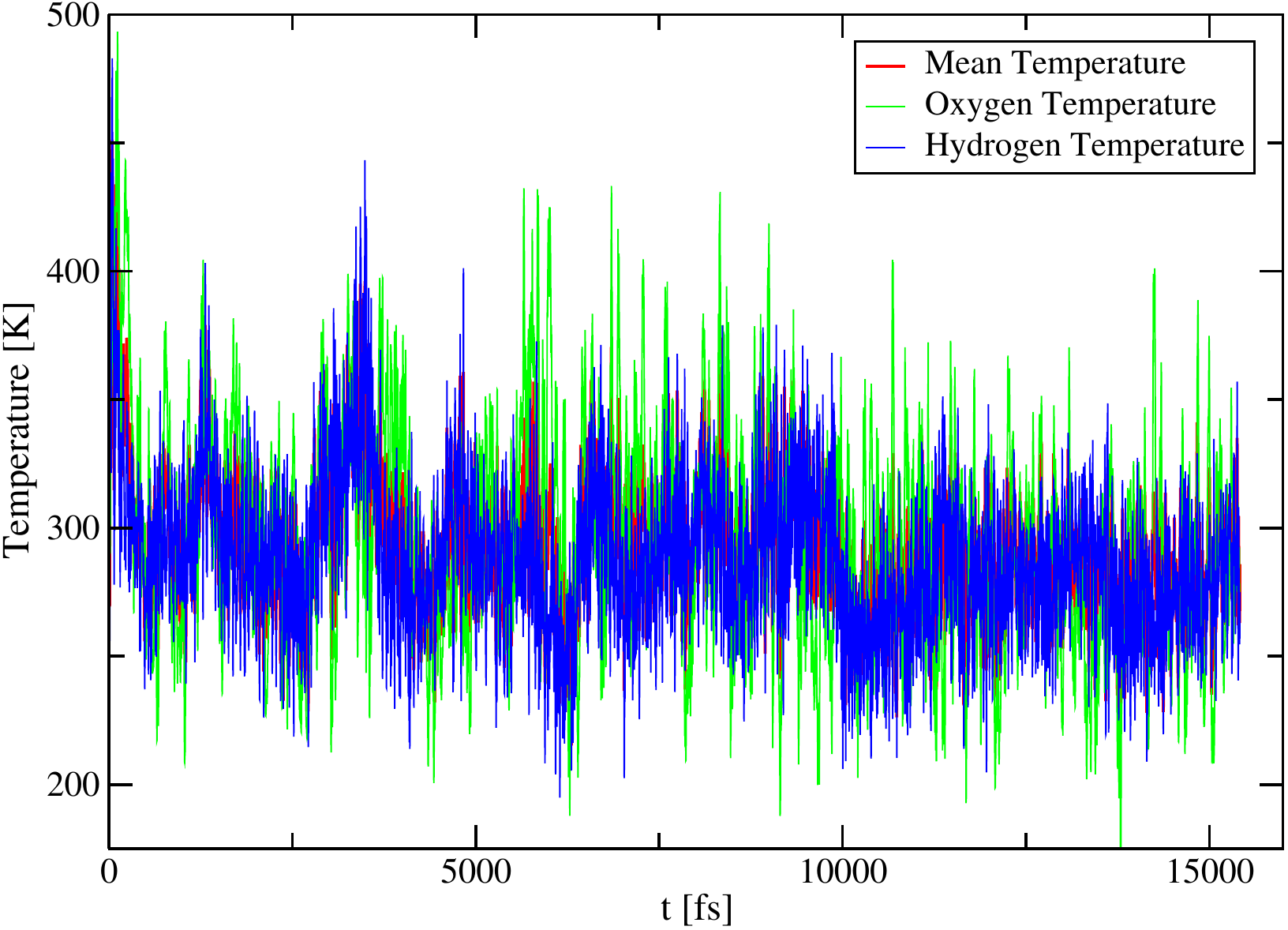}
  \caption{Species-resolved temperature evolution with overlay friction $\gamma_L=3.75\times10^{-4}$~fs$^{-1}$.  The larger overlay noise provides a more conservative canonical thermostat while remaining weak on the scale of the molecular relaxation time.}
  \label{fig:gamma-l-3750}
\end{figure*}

\subsection{Species-dependent friction values}

\begin{figure*}[t]
  \cpincludegraphics[width=\linewidth]{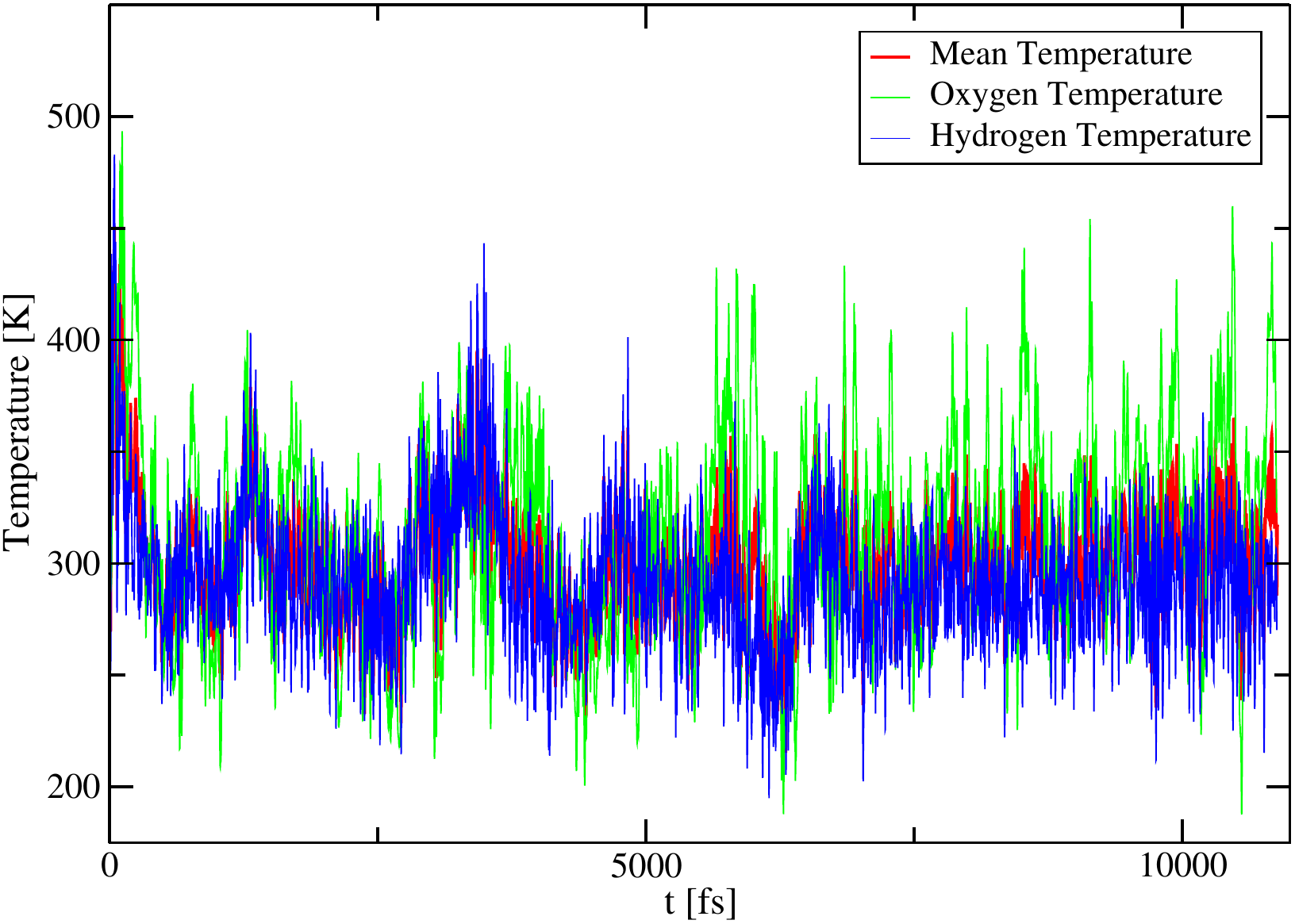}
  \caption{Species-dependent dissipative correction using the empirical mass scaling $\gamma_D^I=\gamma_D\sqrt{\min_I M_I}/\sqrt{M_I}$.  The result is consistent with the expected temperature distribution, but for this water benchmark the global correction is already sufficient.}
  \label{fig:mass-scaled-gamma}
\end{figure*}

For most applications, a carefully chosen global \texttt{NOISY\_GAMMA} combined with a modest overlay \texttt{GAMMA} is sufficient.  If the species-resolved temperatures remain imbalanced, one can add species-dependent corrections with \texttt{GAMMA\_D}.  These corrections may be used either in addition to, or instead of, the global \texttt{NOISY\_GAMMA}.  For example:

\begin{verbatim}
  &MD
    ENSEMBLE LANGEVIN
    GAMMA 0.0
    NOISY_GAMMA 0.00005000
    GAMMA_D  -0.00003125 0.00002500
  &END MD
\end{verbatim}

In this example, \texttt{GAMMA\_D} is interpreted as a correction to \texttt{NOISY\_GAMMA}.  If \texttt{NOISY\_GAMMA} is omitted, the global value defaults to $0$~fs$^{-1}$ and the list supplied to \texttt{GAMMA\_D} gives the species-dependent friction values directly.  The order of the entries follows the order of the atomic species in the \texttt{\&COORD} section.  For water, the example corresponds to $7.5\times10^{-5}$~fs$^{-1}$ for hydrogen and $1.875\times10^{-5}$~fs$^{-1}$ for oxygen, testing the empirical scaling $\gamma_D^I=\gamma_D\sqrt{\min_I M_I}/\sqrt{M_I}$.  Figure~\ref{fig:mass-scaled-gamma} shows that this scaling is plausible, but not necessary for the present benchmark because the global correction already samples the correct temperature distribution.

\subsection{Concurrent relaxation of nuclear and electronic degrees of freedom}

Finally, \cpk{} can perform a ``Langevin annealing'' run in which damped Langevin dynamics relaxes the nuclear and electronic degrees of freedom concurrently.  This procedure is inspired by the original Car--Parrinello idea and can simultaneously relax the structure, improve diagonalization, and restore self-consistency.

\begin{verbatim}
  &MD
    ENSEMBLE LANGEVIN
    GAMMA 0.001
    NOISY_GAMMA 0.00005
    TEMPERATURE_ANNEALING 0.9999975
  &END MD
\end{verbatim}

\begin{figure*}[t]
  \cpincludegraphics[width=\linewidth]{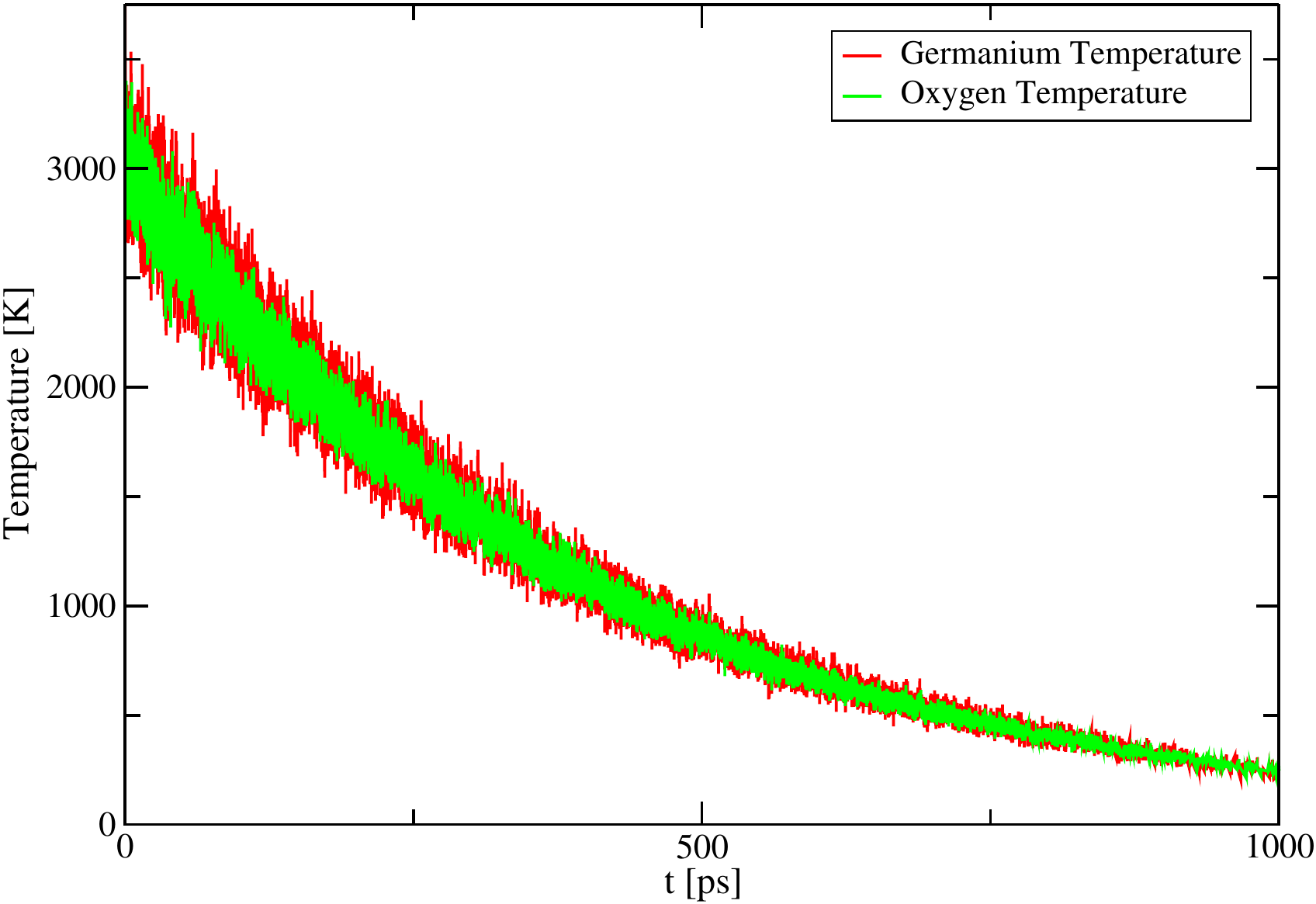}
  \caption{Langevin annealing from liquid to vitreous germania.  The example illustrates concurrent damping of nuclear and electronic degrees of freedom and is included as a practical reference for relaxation workflows beyond the liquid-water benchmark.}
  \label{fig:langevin-annealing}
\end{figure*}

True annealing runs, as opposed to simple quenches, can be expensive enough to dominate the computational budget.  We therefore do not add a separate water annealing example here.  Instead, Figure~\ref{fig:langevin-annealing} illustrates the same capability for the \textit{ab initio} annealing of liquid germania to vitreous germania.

\begin{acknowledgments}
We would like to thank the whole CP2K team, in particular M. Krack, F. Mohamed, M. Iannuzzi and J. Hutter. The generous allocation of computer time from CSCS Manno and the ICT Services of ETH Zurich is acknowledged, as well as corresponding support from Neil Stringfellow and Olivier Byrde, respectively.
\end{acknowledgments}

\appendix

\section{GTH pseudopotentials and Gaussian basis sets}

\begin{widetext}
\begin{Verbatim}[fontsize=\scriptsize]
H GTH-PBE-q1
    1
     0.20000000    2    -4.17890044     0.72446331
    0
#
O GTH-PBE-q6
    2    4
     0.24455430    2   -16.66721480     2.48731132
    2
     0.22095592    1    18.33745811
     0.21133247    0
\end{Verbatim}
\end{widetext}

\begin{widetext}
\begin{Verbatim}[fontsize=\scriptsize]
H DZVP-GTH
  2
  1  0  0  4  2
        8.3744350009  -0.0283380461   0.0000000000
        1.8058681460  -0.1333810052   0.0000000000
        0.4852528328  -0.3995676063   0.0000000000
        0.1658236932  -0.5531027541   1.0000000000
  2  1  1  1  1
        0.7270000000   1.0000000000
#
O DZVP-GTH
  2
  2  0  1  4  2  2
          8.85980961     0.13629371     0.00000000    -0.08866335     0.00000000
          2.79327113     0.02080784     0.00000000    -0.26937441     0.00000000
          0.90727943    -0.60919527     0.00000000    -0.45939673     0.00000000
          0.28741531    -0.50259053     1.00000000    -0.41039240     1.00000000
  3  2  2  1  1
        1.1850000000   1.0000000000
\end{Verbatim}
\end{widetext}

\end{document}